# Analysis of thermal radiation in ion traps for optical frequency standards

M Doležal[1], P Balling[1], P B R Nisbet-Jones[2], S A King[2], J M Jones[2], H A Klein[2], P Gill[2], T Lindvall[3], A E Wallin[3], M Merimaa[3], C Tamm[4], C Sanner[4], N Huntemann[4], N Scharnhorst[4], I D Leroux[4], P O Schmidt[4,5], T Burgermeister[4], T E Mehlstäubler[4] and E Peik[4]

[1] Czech Metrology Institute, V Botanice 4, 150 72 Prague, Czech Republic
[2] National Physical Laboratory (NPL), Hampton Rd, Teddington, TW11 0LW, United Kingdom
[3] VTT Technical Research Centre of Finland Ltd, Centre for Metrology MIKES, P.O. Box 1000, FI-02044, VTT, Finland
[4] Physikalisch-Technische Bundesanstalt, Bundesallee 100, 38116 Braunschweig, Germany
[5] Institut für Quantenoptik, Leibniz Universität Hannover, Welfengarten, 130167 Hannover, Germany

**Abstract.** In many of the high-precision optical frequency standards with trapped atoms or ions that are under development to date, the AC Stark shift induced by thermal radiation leads to a major contribution to the systematic uncertainty. We present an analysis of the inhomogeneous thermal environment experienced by ions in various types of ion traps. Finite element models which allow the determination of the temperature of the trap structure and the temperature of the radiation were developed for 5 ion trap designs, including operational traps at PTB and NPL and further optimized designs. Models were refined based on comparison with infrared camera measurement until an agreement of better than 10% of the measured temperature rise at critical test points was reached. The effective temperature rises of the radiation seen by the ion range from 0.8 K to 2.1 K at standard working conditions. The corresponding fractional frequency shift uncertainties resulting from the uncertainty in temperature are in the $10^{-18}$ range for optical clocks based on the $Sr^+$ and $Yb^+$ E2 transitions, and even lower for $Yb^+$ E3, $In^+$ and $Al^+$. Issues critical for heating of the trap structure and its predictability were identified and design recommendations developed.

## 1. Introduction

Optical atomic clocks have shown rapid improvement over the last decade, with total systematic uncertainties in the $10^{-17}$ to $10^{-18}$ range becoming increasingly common [1],[2]. For many systems, one of the largest contributions to the systematic uncertainty comes from the blackbody radiation (BBR) shift. This quadratic Stark shift arises from the electric field radiated by surfaces at finite temperatures, which interacts with electric dipole moments in the atom or ion and perturbs the energy levels. The shift is not equal in the ground and excited states of the clock transition, and hence there is a systematic perturbation to the resonance that must be accounted for. The frequency shift $\Delta f$ of a particular transition can be expressed [3] as:

$$\Delta f = -\frac{\Delta\alpha \langle E^2 \rangle}{2h}(1+\eta) \quad (1)$$

where $\langle E^2 \rangle$ is the mean-square electric field associated with the BBR, $h$ is Planck's constant, $\Delta\alpha$ is the differential static polarizability between the ground and excited states, and $\eta$ is a (small) dynamic correction that must be applied due to the overlap of the ambient thermal radiation spectrum with the atomic resonances. If the electronic angular momentum of one or both atomic states is larger than one half, $\Delta\alpha$ is generally not scalar but has a non-zero tensorial component so that $\Delta f$ depends on the polarization of the electric field relative to the quantization axis and on the magnetic quantum number(s). Tensorial shifts in optical clocks are usually cancelled by averaging over orthogonal orientations of the applied static magnetic field or different Zeeman components of the clock transition [4],[5]. Therefore, in the following discussion the degree of polarization of the thermal radiation plays no role, and permits the simplifying assumption that the radiation is unpolarized.

From (1), it can be seen that there are three potential sources of uncertainty in the magnitude of the BBR shift. The first is the uncertainty in the differential static polarizability of the two clock states, which arises from the uncertainty in the theoretical calculations and/or direct experimental determination. The second source of uncertainty is that of the mean-square electric field $\langle E^2 \rangle$, which for a perfect blackbody emitter is determined by Planck's law:

$$\langle E^2 \rangle \propto \int_0^\infty \frac{\nu^3}{e^{h\nu/k_B T}-1} d\nu \propto T^4 \quad (2)$$

where $k_B$ is Boltzmann's constant, $\nu$ is the frequency and $T$ is the temperature of the emitting surface. The third source of uncertainty is in the dynamic correction to the shift, $\eta$, which varies strongly with temperature ($\eta \propto T^2$ to first order) and is also limited by theoretical predictions and/or experimental measurement.

One significant advantage of trapped ion optical clocks is that they generally have a smaller static BBR shift and dynamic correction than those based on neutral atoms [1],[2]. This is primarily due to the blue shifting of the atomic resonances as a result of the tighter binding of the remaining electrons after ionization, thereby increasing the detuning between the thermal radiation and the atomic transitions and producing a smaller $\Delta\alpha$ and dynamic correction $\eta$. As an example, the total BBR shift for a room-temperature single ion clock based on the Yb[+] electric octupole clock transition is around fifty times smaller than that for a Sr neutral atom lattice clock [6],[7] which greatly relaxes the requirements on the determination of the BBR temperature. For all ions that are currently studied for use as high precision optical clocks, the dipole transitions that contribute to the BBR shift are at wavelengths that are much shorter than the peak wavelength of thermal radiation at room temperature. Consequently, the dynamic corrections are small: $|\eta_{(300 \text{ K})}|<0.01$ [2].

A major complication in the determination of the BBR shift for trapped ions is that the strong radio frequency electric fields used to confine the ions will heat the dielectrics that are necessarily incorporated into the structure of an ion trap to electrically insulate various components. This heat generation means it is no longer appropriate to use the room or vacuum chamber temperature to estimate the BBR shift, as the trap structure subtends a large fraction of the solid angle visible to the ion and may be at a very different temperature than the chamber. One way of determining the effective temperature of the thermal radiation would be the use of in situ sensors [8]. This approach is difficult to apply for ion traps due to the strong rf field which will cause heating of the sensor, and any electronics attached to the sensor will experience strong interference unless precautions are taken. The presence of the sensor may also disturb the trapping potential, and therefore it is more common to indirectly estimate the BBR from temperature measurements in dummy setups and characterization of the emissivity ($\varepsilon$) of surfaces [9].

The effective temperature seen by the ion can be estimated by simulation of the trap heating via finite-element method (FEM) modelling. These models calculate the temperature rise of the trap parts and the resulting steady-state temperature distribution due to thermal conduction and radiation. This can then be compared with experimental observations using an infrared (IR) camera. This comparison is essential for validation of the model to account for manufacturing tolerances, and as some material and surface properties are not exactly known.

In this paper we present FEM models for thermal analysis of several different ion trap designs, some of which are presently in use for optical atomic clocks but were designed with less attention paid to thermal behaviour, and others which have been specifically designed to have low levels of heating. We validate these models against experimental observations on dummy traps. Initial discrepancies are analyzed, explained and corrected for. From the models and observations we present values for the effective BBR temperature that would be observed by trapped ions situated in these traps and a resulting uncertainty in the BBR shift.

The structure of the paper is as follows: Section 2 consists of a discussion of the origin of ion trap heating and potential methods to minimize the effects and contains recommendations for future ion trap designs; in section 3 we describe the principles and problems of modelling; in section 4 we briefly present temperature measurements of a dummy trap; and in section 5 we present models and experimental details of five ion traps that are currently in use and discuss possible ways for further decreasing the temperature rise.

## 2. Heat dynamics in the trap

We have identified two sources of heat generation in the trap, rf absorption in insulators and joule heating in conductors, as well as two ways of heat removal from the trap, radiation and conduction.

### 2.1. Heat generation

*2.1.1. Insulators (dielectrics).* The heat power dissipation density per unit volume $p$ in a dielectric located in an ac electric field $E$ is given [9] by:

$$p = 2\pi f \varepsilon_0 \varepsilon_r \tan(\delta)|E|^2 \quad (3)$$

where $f$ is the frequency of the electric field, $\varepsilon_0$ is the permittivity of free space, $\varepsilon_r$ is the real relative permittivity and $\tan(\delta)$ is the dielectric loss tangent. Both $\varepsilon_r$ and $\tan(\delta)$ are functions of the frequency $f$. For many materials used in ion traps, these values are not known for the specific sample and/or operating frequency range. Wherever possible it is therefore recommended to directly measure these values for the material that will be used in the trap. If this is not possible literature values can be used as an initial estimate for FEM modelling and refined, with an appropriate uncertainty, to fit the thermal measurement data. Table 1 summarizes material properties of dielectrics commonly used in ion traps.

Due to manufacturing tolerances, the geometry of the insulator body may differ from the design. If a small amount of material is removed at a place of high heat generation the dissipated power is decreased accordingly. Dielectrics can also influence the heat generation in the trap indirectly by increasing the total capacitance of the trap which results in

Analysis of thermal radiation in ion traps for optical frequency standards, arXiv:1510.05556 [physics]     3higher currents flowing through the conductors, and consequently produces higher losses through joule heating. One example of this indirect influence can be demonstrated in the case of the endcap trap [14] design shown in Figure 1 (a).

The heat generated in the trap is sensitive not only to the dielectric loss tangent, but also to the permittivity (see (3)). For example, alumina has a relatively low loss tangent between $1\times10^{-4}$ and $2\times10^{-4}$, but a high relative permittivity $\varepsilon_r \sim 9.5$. This results in a high capacitance of the cylindrical capacitor created by electrodes with alumina dielectric inside, and consequently higher currents through the leads. However, small inevitable gaps between the insulator and electrodes will act as capacitors in series and thus decrease the total capacitance.

Another flaw in this design is the direct thermal contact between the dielectric and the inner electrodes, which results in heat transfer between the two. As the inner electrodes represent a large solid angle visible to the ion, an increase in their temperature must be avoided.

An improved version of the endcap trap that addresses these problems is shown in figure 1 (b). The rf is carried into the trap via a bulk copper body, through which the inner molybdenum electrodes are sunk. The outer electrodes are mounted to the copper body using fused silica insulators, which have better electrical properties than alumina (table 1).

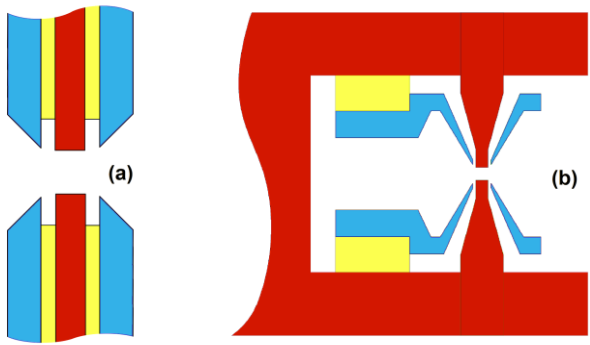

**Figure 1.** Endcap trap electrodes with dielectric insulators as spacers (cross section), (a) - established design, e.g. [14]; (b) - improved design. The rf electrodes are in red, dc electrodes are in blue, and insulators are in yellow.

The capacitance between the inner and outer electrodes is reduced by increasing their separation with a conical geometry. The insulators have been moved to a region where they are not in direct contact with the inner electrodes to reduce heat transfer between the two. The electric field strength across the dielectric is reduced by increasing the dielectric thickness; in spite of the proportional increase of the volume with increasing thickness, the total heat power dissipation decreases due to the quadratic dependence on $E$, see (3).

*The design recommendations are to minimize the trap capacitance through the geometry of the electrodes and to use low loss insulator materials like fused silica or sapphire placed in locations with low electric field strength.*

### 2.2. Conductors

The heat generated in the conductors is the well-known joule heating that depends on the electrical resistivity of the materials. Since the traps are operated at frequencies where the skin depth is important, it may be beneficial to coat the conductors (preferably of high conductivity themselves) with a material of even higher conductivity. A common coating choice is gold which has a conductivity of 41 MS/m and has the additional benefit of a low emissivity. Exceptions exist however. For example, gold plating of copper is not favourable since nickel, which has a high magnetic permeability, is needed as an intermediate layer. This affects both the skin depth and the stringent requirements for controlling the magnetic fields in the trap.

**Table 1.** Selected material properties of dielectrics used in traps discussed in this work (*rows printed in italics correspond to interesting alternatives*).

| Material | Conductivity (S m$^{-1}$) | Thermal conductivity (W m$^{-1}$K$^{-1}$) | Emissivity (7-13) µm[a] polished | Emissivity (7-13) µm[a] matt | Relative permittivity | Dielectric Loss tangent ($10^{-4}$) |
|---|---|---|---|---|---|---|
| Shapal | $<1\times10^{-10}$ | 90 | | 0.867 | 7.1 | 10 |
| AlN ceramics | $<1\times10^{-12}$ | 140...180[c] | 0.73 | 0.78 | 9 | 3[c]; 5[c] |
| Sapphire | $<1\times10^{-16}$ | ⊥ 30.3; ∥ 32.5 | 0.788 | | 9.3...11.5 | 0.2 |
| Macor | $<1\times10^{-14}$ | 1.46 | | 0.919 | 5.67 | 11[c]; 20[c] |
| Alumina cer. | $<1\times10^{-12}$ | 4...35[c] | | 0.855 | 9.5 | 2[c] |
| Fused silica | $<1\times10^{-18}$ | 1.35 | 0.746 | | 3.82 | 0.15 |
| *Diamond* | *-* | *900...2300...3320[c]* | *< 0.03[b]* | | *5.7* | *0.2* |
| *BeO* | *$<1\times10^{-13}$* | *285* | | | *6.76* | *4* |

[a] The total emissivity may differ from spectrally limited values shown in this table. The values shown here were measured for samples of materials used in dummy and operational traps and are used for corrections of raw IR camera data acquired with the same type of camera and vacuum window. For example, the total emissivity of sapphire at room temperature is just 0.47...0.57 [11] but effective value (measured in this work in above mentioned spectral band) is 0.788 due to the strong spectral dependences of absorption and reflectivity of sapphire [12].

[b] Pure diamond is transparent in an extremely broad spectral range, so its emissivity is low. The value shown here is rather an upper limit measured at elevated temperature [13].

[c] Range of values can be found in literature or datasheets (values for ceramics strongly depend on composition, processing and purity). Values used in FEM models are discussed in the text.

The skin depth $\delta$ can be expressed by:

$$\delta = \sqrt{\frac{2\rho}{2\pi f \, \mu_r \mu_o}} \qquad (4)$$

where $\rho$ is the electrical resistivity of the material, $f$ is the frequency of the field, $\mu_r$ is the relative permeability and $\mu_0$ is the permeability of free space.

Calculating the skin depth for the materials intended for use and for the planned frequency may help to estimate the necessary thickness of coating or the size (diameter) of conductors needed. The skin depth in conductor materials commonly used in ion traps is of the order of (15...50) µm ($f \sim 20$ MHz).

*The design recommendations are to minimize heat generation in conductors by using materials with high conductivity (see table 2) and consider an additional coating, thickness and shape with respect to the skin effect (e.g. sheet metal instead of wires).*

### 2.3. Efficient heat removal

*2.3.1. Radiation.* Radiative heat dissipation from the trap is problematic since this radiation will be seen by the ion and give rise to BBR shifts. Decreasing the emissivity, and thus thermal emission of the hot components (especially those in the field of view of the ion), decreases the temperature rise seen by the ion by an amount corresponding to the change in emissivity and the solid angle fraction occupied by that component. This can be clearly seen in Figure 2. Due to the large solid angle that the faces of the endcap (inner) electrodes occupy, their radiative emissions will dominate the thermal field to which the ion is exposed.

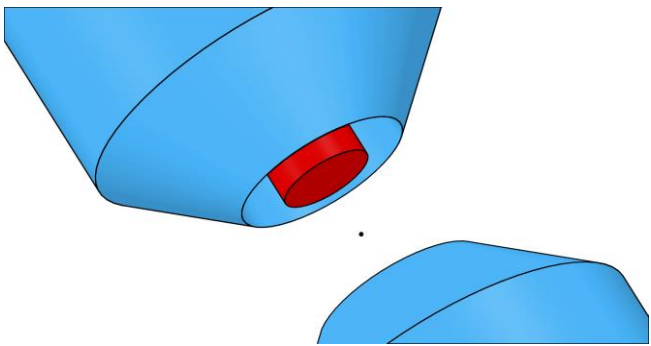

**Figure 2.** Endcap design with illustration of the ion position (small sphere - not to scale). Coating or polishing of all surfaces may be advantageous but the circular faces of inner electrodes deserve the highest priority since they occupy the largest part of the field of view from the ion location.

The emissivity of any given component is determined by a combination of the material that it is composed of and its surface quality. Surface roughness generally increases the emissivity, while polishing will decrease it down to a minimal value which depends upon the material (highly polished silver 0.01, copper and gold 0.02, molybdenum 0.02 [15], tantalum 0.17[6]). Whilst metals tend to have a reasonably flat spectral dependence in the IR the emissivity of dielectrics is often highly spectrally dependent, with many insulator materials (Macor, alumina, glasses, and sapphire) being "white" (or transparent) at visible wavelengths, but "black" (or absorbing) in the IR spectral range relevant for blackbody radiation. For materials with a high intrinsic emissivity, covering the pre-polished electrode materials by a thin electroplated / sputtered gold or silver layer can effectively minimize the emission. Whilst using materials with a low emissivity can reduce their radiative effect, it is recommended to place hot insulators outside of the field of view of the ion. This can be done either by shielding them with a cold black shield or by a reflecting shield which directs the radiation from the hot emissive part away from the ion. Even with low emissivity materials, gaps or holes appear to have a higher emissivity than a flat surface due to the multiple internal reflections. If at all possible all gaps and internal surfaces which might be at a raised temperature should be completely out of the ion's direct line of sight.

**Table 2.** FEM parameters used for conductors. The empty brackets (..) indicate that this table entry was not required for the calculation.

| Material | Electrical conductivity (MS m$^{-1}$) | Thermal conductivity (W m$^{-1}$ K$^{-1}$) | Emissivity [b] |
|---|---|---|---|
| Copper | 58.00 | 400 | 0.10 |
| Molybdenum | 17.60 | 139 | 0.162 |
| Titanium | 1.82 | 21 | 0.166 |
| BeCu | 31.90 | 120 | 0.184 |
| 304 Stainless steel | 1.39 | 16.2 | (..) |
| Mumetal | 1.72 | 34.6 | (..) |
| Tantalum | 6.30 | 57.5 | 0.30 |
| Tantalum polished | 6.30 | 57.5 | 0.15 |
| Stainless steel | 1.10 | 15.1 | 0.25 |
| Aluminum | 36.00 | 150 | 0.25 |
| AlMgSi0.5 | 30.00 | 210 | 0.166 |
| Ti6Al4V | 0.56 | 6.7 | 0.166 |
| Gold (sputtered) [a] | 26.00 | 315 | 0.05 |

[a] The electrical conductivity of sputtered gold layer on AlN was measured and found to be lower than 41 MS/m typical for pure solid gold.
[b] The emissivity was estimated by IR camera measurement for samples with the same or similar surface treatment as used in ion traps. The repeatability of these measurements was about 1%.

The above measures aim at decreasing the emissivity of hot parts, however it is also beneficial to increase the emissivity (absorption) of cold parts. A significant part of the solid angle seen by the ion is occupied by the inner surface of the vacuum chamber and windows. If these are highly emissive they will not reflect the radiation from hot parts of the setup, but emit well defined radiation corresponding to ambient temperature. Vacuum windows typically have quite high emissivity in the IR so they reduce the temperature rise from radiation seen by the ion more efficiently than typically reflective metal walls of the chamber.

---
[6] The polished molybdenum emissivity estimated in CMI by a camera sensitive in the (7-13) µm region resulted in a higher effective emissivity value of (16.2±3.4)%.

*The design recommendation is to minimize the emissivity of hot parts which cover a high portion of the solid angle visible from the ion location. This can be done by polishing or coating. It is also advisable to increase the emissivity of cold parts. In particular, the emissivity of the inner vacuum chamber walls could be increased by coatings[7] or simply by roughening on the length scale of the BBR wavelength range.*

### 2.3.2. Heat Conduction.
Conduction is the most important mechanism of heat disposal. For the traps where the components only see a small temperature rise (like the new NPL endcap trap design described in section 5.3.1), conduction dominates the heat removal from the trapping region. One must take care however to ensure that externally generated heat, for example from the vacuum feedthrough insulator, is not conducted to the trap but is disposed of in another way (e.g., active feedthrough cooling, "thermal grounding" to an efficient heat sink or selection of a feedthrough insulator material with low heat generation).

*The design recommendation is to use good heat conductors with large cross section (see table 1 for dielectrics and table 2 for electrical conductors). Special attention should be paid to establish good thermal contacts between the parts, since these very often appear to be decisive for the heat dynamics in the trap and consequently its temperature rise above the ambient temperature (see section 2.4). For the rf vacuum feedthrough, it is important to pay attention to the selection of an insulator with low loss tangent, or consider cooling (as is common in rf power feedthroughs).*

## 2.4. Special considerations

### 2.4.1. Thermal contacts.
The important role of thermal conductivity at the contact of two materials was identified during the experiments with dummy traps. The thermal contact conductivities are typically lower in vacuum than in air where the thermal conduction of residual gas can be neglected [17]. For example, a stainless steel to stainless steel contact [18] has a thermal conductance between (2000 and 3000) W m$^{-2}$ K$^{-1}$ in air and between (200 and 1100) W m$^{-2}$ K$^{-1}$ for evacuated gaps. The roughness of the surfaces in contact and the pressure that holds the parts together is also of great importance. We indirectly observed different thermal contact conductance (typically between a conductor and an insulator) when repeatedly disassembling and re-assembling the dummy traps, which resulted in different temperature gradients across the contacts. The temperature gradients varied from one assembly of the trap to another over a range of approximately 20%[8]. To verify that the thermal gradients were caused by a varying thermal contact conductance, a small amount of heat sink compound was applied between the surfaces and the gradients across the contacts dropped (conductive silver heat sink compound Primecooler PC-C100S and electrically insulating silicone heat sink compound Arctic Cooling Silicone was used). We found that the added compound did not change the voltages at the trap electrodes nor did it introduce additional electrical losses. The heat sink compound used is not UHV compatible, but UHV compatible thermal contact pads or liquids do exist.

*The design recommendation is to ensure good heat transfer through mechanical contacts by well matching surfaces, preferably with polished contact faces, or to consider soldering, gluing, inserting soft metal foil (indium) or plating with soft metal (e.g. silver).*

### 2.4.2. Feedthrough.
The thermal properties of the rf vacuum feedthrough deserve significant attention, especially when the trap is mounted directly onto the feedthrough pin. In this situation as the feedthrough serves as a heat sink, its self-heating should be minimized and the atmosphere-side part of the feedthrough should be cooled or temperature stabilized.

The electrical feedthrough pins on which the traps are mounted can be made from different materials. Stainless steel pins are commonly used, however they are not the best option as stainless steel has a relatively low thermal and electrical conductivity and the pins themselves are hollow in some cases. The result of this is that the temperatures of the trap parts strongly depend on the trap location on the pin (this was one of the underestimated parameters affecting repeatability during the measurement of dummy traps – e.g. for 100 mW heat transfer the temperature gradient along a stainless steel pin with a cross section of 1 mm$^2$ is approximately 6.6 K per millimeter of the pin length). Instead, the pin/rod should be thick, solid, and made of a material with high thermal conductivity such as copper.

The second important topic that has to be considered when selecting the electrical feedthrough is the insulator. The dielectric losses (and therefore the heat generated) in the feedthrough insulator could be comparable to the heat generated in the trap. The heat generated in the insulator material depends on the electric field which exists in the bulk of the insulator. The electric field and consequently the heat generation are lower if the insulator diameter is larger. Another option to decrease the heat generation (for the same insulator volume) is to use a custom feedthrough where the ground conductor is not attached to the insulator surface along the whole insulator (see Figure 3).

---

[7] For example coating by non-evaporable getter or by carbon nanotubes [16]

[8] Even reducing the roughness of the parts by repeatedly disassembling and re-assembling the trap produced a noticeable improvement of the thermal contacts. We observed smaller and smaller temperature gradients after each reassembly.

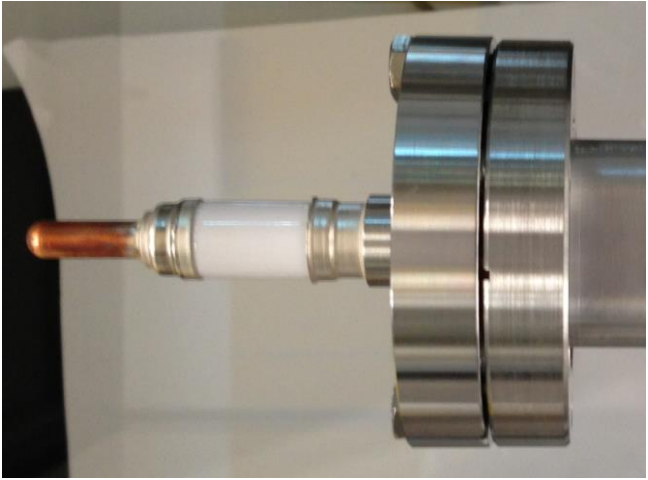

**Figure 3.** Detail of vacuum feedthrough with visible "open" insulator and thick copper rod.

*The design recommendation is to use an electrical feedthrough with thick highly conductive (e.g. copper) pins and to minimize the shielded area of the insulator.*

*2.4.3. Wires & Strips.* The conductor (copper) strips shown in Figure 4 are one of the design patterns used in ion traps. As a result of the skin effect, copper strips have significantly improved electrical conductivity over round wires that have the same cross-section area. The nonzero skin depth and the thermal conductance requirements impose conditions on the minimum thickness of the strips. From electrical considerations, a thickness of 50 µm is sufficient for a skin depth of 15 µm at a trap drive frequency of 20 MHz, whereas FEM simulations show that the temperature rise of the strips can be substantially reduced through enhancement of the heat conductivity by increasing the thickness to e.g. 200 µm.

*The design recommendation is to use strips as thick as their stiffness allows.*

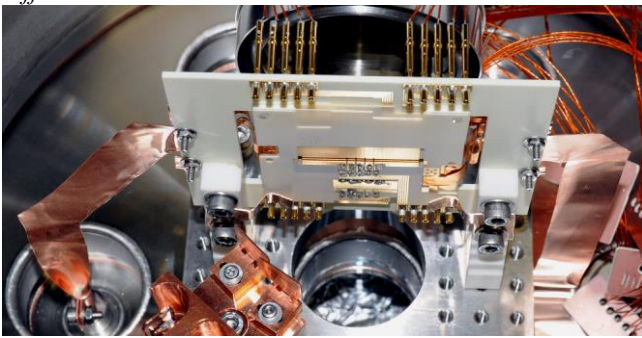

**Figure 4.** Trap assembly for PTB In+/Yb+ multiple ion clock.

*2.4.4. Chamber and ambient environment.* External sources of heat (such as the rf source including helical resonator and feedthrough, and also the vacuum pump) should be cooled separately and the temperature of the vacuum chamber should be measured at several points as gradients comparable to the target uncertainty on the order of 0.1 K are likely to be present.

## 3. FEM analysis

FEM analysis is a tool that helps to estimate the effective temperature of the radiation seen by the ion. Unfortunately we have not found an ideal FEM software package on the market. Both leading software brands COMSOL and ANSYS use a similar implementation of radiation in their "thermal" solver [19].

*3.1. Software limitations*

In ANSYS, bodies are represented as grey opaque with diffusive surfaces. This means that neither specular reflections nor the wavelength dependence of the emissivity can be captured in the FEM model. Neither limitation is expected to significantly increase the uncertainty of our calculations because the critical specular reflections are not produced by strongly curved (convex) surfaces (trap electrodes and mounts) and the employed materials show only little emissivity variations in the central range of the BBR spectrum.

The other issue is that the space between model bodies is not meshed and hence we cannot deduce the temperature of the radiation there. To get this information from the FEM model it is necessary to include a new body located at the position of the ion, whose calculated temperature serves as an estimate of the effective temperature seen by the ion. The body is a very small blackbody sphere with unity emissivity, to prevent the sphere from influencing the radiation in the trap due to reflection, and with a thermal conductivity of 1000 W m$^{-1}$ K$^{-1}$ that is high enough to result in a negligible temperature inhomogeneity. We tested the models with different sphere diameters (from 1 µm to 80 µm) and the results were the same within ±0.01 K. These discrepancies were effects of the re-calculated FEM mesh rather than the sphere size. We can conclude that the sphere size does not influence the results. This approach is suitable for an evaluation of the BBR shift if the temperature differences between different surfaces seen by the ion are much smaller than the average ambient temperature (in Kelvin), and if the dynamic correction $\eta$ (Eq. (1)) is small [8],[20]. Both criteria are well fulfilled in this work − temperature rise of surfaces is in all cases lower than 10 K.

*3.2. Simulation setup*

The trap heating has been modelled using a two-step process. First we used the ANSYS HFSS component to solve the electromagnetic part of the model. This part calculates losses (heat generation) in the trap parts based on the geometry, material parameters and applied voltage. For the heavy duty conductors the calculation was done also in the bulk (skin effect captured by dense meshing) but for the vacuum chamber and the flanges a simplified approach was used and we calculated currents as if they were surface currents (the skin effect is not resolved by meshing but the skin effect resistance is estimated from analytical formulae). The second part of the calculation was the determination of heat dissipation through the trap by conduction and radiation (face-to-face among appropriate surfaces). The heat generation calculated in the first part was used as the load for the "ANSYS Steady state thermal" component that calculated the steady-state temperature map of the trap. The ambient temperature was set to 20°C in all simulations, which is applicable to the temperature map pictures presented below. To get the temperature rise of the trap parts the 20°C ambient temperature has to be subtracted from the values presented in the figures.

### 3.3. Simulation optimization

Experimental imperfections were not included in the initial FEM simulations which caused the calculated temperature maps to differ significantly from the observed temperature rises. In this situation, the simulations and experimental observations were amended in order to reveal the origin of the discrepancies. Typically the material parameters are not known at the rf frequencies at which the traps operate, and the emissivity of the trap materials is not published or covers a wide range of values. Moreover, there may be problems with manufacturing tolerances (such as gaps between trap parts influencing the trap capacitance and consequently heat generation) or unknown thermal contact conductivities, dielectric loss tangents etc. Great care must be taken when modifying the model to match the observations as the cause of any deviation between model and experiment can have multiple origins and several separated experiments may be needed for their identification.

### 3.4. General results of FEM

The dependence of the temperature rise from the radiation seen by the ion $\Delta T_{ion}$ on the rf voltage amplitudes $U$ can be well approximated by a quadratic dependence

$$\Delta T_{ion} = a\, U^2 \qquad (5)$$

where $a$ is a constant.

As the heat generation in both conductors and dielectrics is proportional to $U^2$, the temperature increase of the trap parts will, to a first approximation, be proportional to $U^2$ as well. Thermal radiation depends on the 4$^{th}$ power of the temperature and causes deviations from expression (5), but both model and experiment show that these deviations are negligible for relatively low temperature rises where radiative heat removal does not play a dominant role (as in all cases discussed in section 5). We did not observe any deviation from a quadratic dependence larger than experimental uncertainty (~0.1 K + 2% of measured temperature rise).

### 4. Dummy trap temperature measurement

An infrared (7-13) μm camera (FLIR A615) measurement was used for validation of the FEM models. The thermal image measurement is especially useful for estimation of the temperature of the insulators. Insulators typically have high emissivity (>0.5) so the temperatures are easy to measure with the camera, but less convenient to measure using sensors due to the high electric field (which induces rf heating of the sensor and causes interference) and the low thermal conductivity (temperature gradient between sensor and material). The IR camera also helps to identify the sources of heat from the rate of temperature rise after switching the rf on due to its fast measurement rate (hundreds of thousands of measurement points/pixels updated several times per second). Temperature sensors are used for measurement of the vacuum chamber's temperature at several points, and also of that of the rf feedthrough. In some cases we have used Pt100 sensors and thermistors for estimation of trap temperatures in vacuum; the heating of the sensors by the rf field was reduced by shielding of the leads and using bypass capacitors to short the rf pickup.

The IR camera sensitivity, vacuum window transmission and the emissivity of material samples used in the traps (metals and insulators of different surface finish) were calibrated using a thermostat (3°C to 50°C) with a nearly perfect emitter (a blackened hole in a copper block) as a reference. In order to avoid convective heat transfer and to keep the sample temperature as close to the thermostat temperature as possible, the thermostat assembly was in vacuum. Even then, temperature gradients due to radiative heat transfer had to be accounted for in the case of samples with low thermal conductivity and/or significant thickness. The vacuum also prevents possible problems with water condensation at surfaces. Having the chamber at a known temperature improves characterisation of background ambient radiation. The uncertainty of temperature estimation by the IR camera is increased for low emitting materials (like polished metals) so in some cases we applied some black thin (0.1 mm) tape with calibrated emissivity (95%) for enhancement of the resolution of temperature gradients in that region. The heating of the tape by the rf field was estimated to be 0.1 K, whilst not negligible the extra heating this introduced is acceptable compared to other uncertainties.

The dummy trap temperature measurements were performed in high vacuum (< $10^{-3}$ Pa) and the operating traps are kept under ultra-high vacuum (< $10^{-8}$ Pa), so the heat transfer by conduction of residual gas can be neglected.

### 5. Analysis of selected ion traps

In following sections we describe the thermal analyses of three designs of 3D RF Traps and of two designs of Linear RF Traps. In each case it contains a trap description, a table with material parameters, a figure of the modelled temperature distribution, the temperature of the BBR seen by the ion with discussion, the uncertainty of the BBR shift resulting from the uncertainty of the BBR temperature and a heat balance table. The small discrepancies in heat balance tables presented in section 5 (total RF heating is not exactly equal to total radiation plus total conduction) are due to the numerical precision of the FEM simulation caused by limiting the mesh size and solver settings to keep computation time within 1-2 days. Uncertainties (95 %) of the temperature rises seen by the ions stated in following subsections were estimated using sensitivity coefficients calculated by FEM for selected input parameters. The uncertainty of the input parameters was based on experience with dummy and operational traps and was conservatively estimated in cases where compelling evidence was not available. In each case we have varied the emissivity of surfaces (trap structure, windows and chamber), loss tangent of dielectrics, input voltage, and the vacuum chamber's temperature homogeneity and uncertainty. In the case of traps mounted on feedthrough we also examined the temperature of the feedthrough (on the accessible atmospheric side) and the thermal conductivity of the feedthrough-trap contacts. The four most important uncertainty contributions are listed in result section of each trap.

## 5.1. PTB Paul trap.

*5.1.1. Trap description.* A cylindrically symmetric Paul trap with conical electrodes is employed for PTB's optical frequency standards based on the electric quadrupole (E2) and the electric octupole (E3) transitions of $^{171}$Yb$^+$ [7],[21]. The electrode shapes were chosen according to a calculation of Beaty (case 4 in table I of [22]) in order to minimize higher-order spherical harmonics of the trap potential. The design is scaled to obtain an endcap electrode diameter of 1 mm. The electrodes are machined from molybdenum. Macor rings are used as insulating spacers between the electrodes. These elements are clamped together by a titanium bolt and nut. The trap mount and the wires leading to the endcap electrodes are connected to a Macor support ring that is attached to the stainless-steel part of the vacuum chamber by three 150 mm long titanium rods (see Figure 5). The trap mount is housed in an extension of the vacuum chamber made from fused silica (see Figure 6).

The radial secular frequency $\omega_r$ of a single ytterbium ion for rf amplitude 600 V at a frequency of 15.56 MHz is $\omega_{r(171Yb+)} = 2\pi \times 0.6$ MHz in this trap.

*5.1.2. Results and discussion.* We measured the temperature distribution in a dummy version of the trap. This dummy trap is made from the same materials as the original trap but is mounted in a compact stainless-steel vacuum chamber where it is held by an electrical feedthrough without further mechanical support.

The temperature distribution among the components of the dummy trap was measured several times after disassembly and reassembly. This included measurements after application of thermally conductive compounds on critical surfaces and after replacing the Macor insulation rings with Shapal [23] rings. These measurements served as a basis for the FEM modelling of the operational PTB trap system that has a more complex vacuum and mechanical support setup as described above.

The material parameters used in that simulation are listed in table 2 and table 3, the resulting temperature distribution is presented in Figure 5 and Figure 6 and corresponding heat balance in table 4.

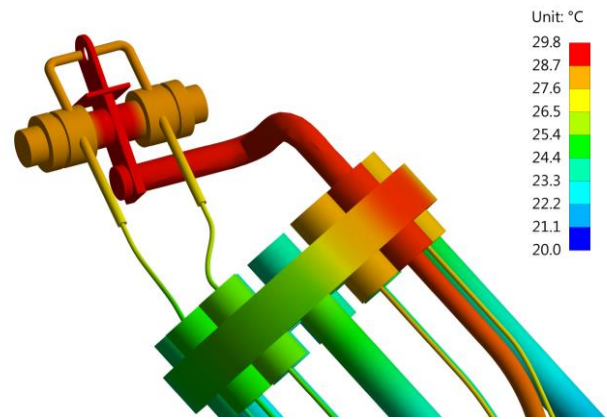

**Figure 5.** Modelled temperature distribution in the operational PTB Paul trap for a 15 MHz rf trap drive voltage of 600 V amplitude applied to the ring electrode. The FEM simulation also included the fused-silica section of the vacuum chamber (see Figure 6) and the electrical vacuum feedthroughs.

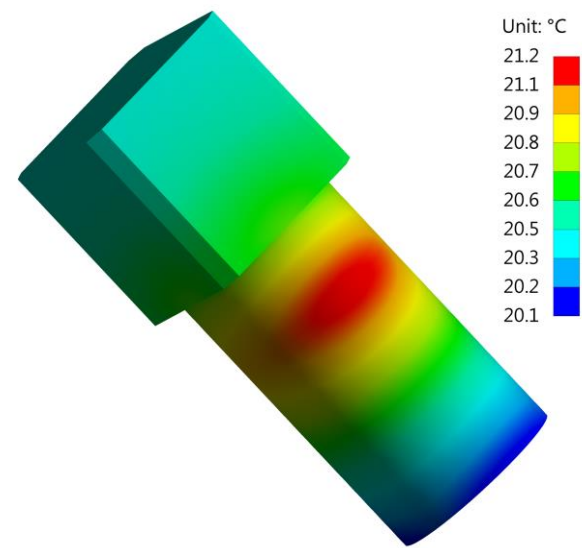

**Figure 6.** Modelled temperature distribution at the outer surface of the fused-silica section of the vacuum chamber of the operational PTB Paul trap. The highest temperature appears close to the Macor support ring segment that is most strongly heated by dielectric loss (see Figure 5).

The data in table 4 provides information on the main sources of rf heating and the main thermal dissipation paths in the operational trap setup. It is worth noting that the largest heat source is a segment of the Macor ring that holds the trap electrode system and is located away from the trapping region. This result of the FEM simulation was confirmed by a direct measurement of the temperature of the outer surface of the fused-silica vacuum chamber section (red spot in Figure 6): here the measured and calculated temperature rises agree within 0.1 K.

**Table 3.** Material parameters used in FEM simulation of the PTB Paul trap. The empty brackets (..) indicate that this table entry was not required for the calculation.

| Item<br>Material | Insulators<br>(in trap)<br>Macor | Feed-through<br>insulator<br>Alumina | Vacuum<br>chamber<br>Fused silica |
|---|---|---|---|
| Relative permittivity | 5.67 | 9.5 | 3.82 |
| Dielectric loss tangent | $1.1\times10^{-3}$ | $2.0\times10^{-4}$ | $1.5\times10^{-5}$ |
| Thermal conductivity (W/m$^{-1}$·K$^{-1}$) | 1.46 | 35.4 | 1.35 |
| Emissivity | 0.92 | (..) | 0.75 |

As indicated in Figure 5, the calculated temperature rise of the hottest part of the trap setup (Macor insulators) is 9.8 K under standard operating conditions. The effective temperature rise of the BBR incident on the ion is calculated as $\Delta T_{\text{ion}(600V, \text{macor})} = (2.1\pm1.3)$ K.

In spite of the higher quality components used in the operational trap and negligible thermal gradients across the contacts between insulator and electrodes in dummy trap and in spite of the fact that measured temperature rise on the outer surface (red spot in figure 6) and the temperature rise predicted by the FEM model without increased thermal resistance of contact agree well, we consider it prudent to include possible gradient due to loose thermal contact at the feedthrough observed in dummy trap also to the uncertainty of the operational trap.

**Table 4.** Heat balance table for PTB Paul trap for drive voltage amplitude 600 V @ 15 MHz.

|  | Part | Power (mW) |
|---|---|---|
| RF Heating | Macor trap electrode spacer rings | 55.9 |
|  | Macor support ring | 114.7 |
|  | RF conductor (Cu rod and wire) | 7.2 |
|  | other components in vacuum chamber | 5.8 |
|  | **Total heat generation** | **183.7** |
| Radiation | Macor trap electrode spacer rings | 4.5 |
|  | Macor support ring | 34.2 |
|  | RF conductor (Cu rod and wire) | 3.8 |
|  | other components in vacuum chamber | 13.8 |
|  | **Total radiation** | **56.4** |
| Conduction | Ti rods | 72.8 |
|  | RF conductor (Cu rod and wire) | 33.1 |
|  | other wires | 21.9 |
|  | **Total conduction** | **127.8** |

The four most important contributions to the uncertainty are the Macor loss tangent (0.93 K), emissivity of molybdenum (0.47 K), voltage amplitude (0.45 K) and uncertainty of thermal conductivity of contacts (0.42 K).

The assumption of such a large relative uncertainty takes into account the possibility of significant batch-to-batch variations of the Macor loss tangent ($5\times10^{-4}$ corresponding to relative uncertainty of ~50%). Such variations became obvious when we compared temperature measurements of the dummy traps manufactured at PTB and NPL (see sections 5.1 and 5.2) with our FEM simulations.

The calculated uncertainty of the effective temperature of the BBR present in the PTB Paul trap leads to uncertainty contributions to the BBR shifts of the E2 and E3 transitions of $^{171}$Yb$^+$. With the established data for the differential polarizabilities of the Yb$^+$ reference transitions [7],[24], we obtain fractional uncertainty contributions of $(d\nu/\nu)_{E2} \sim 1\times10^{-17}$ for the E2 and $(d\nu/\nu)_{E3} \sim 2\times10^{-18}$ for the E3 transition frequency for the simulated temperature uncertainty. Presently these uncertainty contributions are much smaller than those resulting from the uncertainties of the differential atomic polarizabilities (see (1)). The temperature uncertainty however will become important if improved experimental determinations of the differential polarizabilities of the Yb$^+$ reference transitions or atomic-structure calculations become available.

### 5.2. Schrama design NPL endcap trap

*5.2.1. Trap description.* Endcap traps based on the Schrama design [14] have been used at NPL for many years for single-ion optical clocks in $^{88}$Sr$^+$ [25] and $^{171}$Yb$^+$ [26] reaching systematic uncertainties at the $10^{-17}$ level. The trap is formed of concentric dc (ID=1 mm) and rf (OD=0.5 mm) tantalum electrodes separated by a 250 μm thickness alumina tube, held in place with an interference fit. A cross section of the trap structure is shown in Figure 1(a). The trap electrodes are aligned and supported by a C-shaped Macor holder, which is then mounted onto a tantalum wire vacuum feedthrough. Electrical and thermal connections between the trap electrodes and the feedthrough are achieved through spot-welding.

The radial secular frequency $\omega_r$ of a single ytterbium ion for rf-amplitude 150 V at a frequency of 14.55 MHz is $\omega_{r(171Yb+)} = 2\pi\times1$ MHz in this trap.

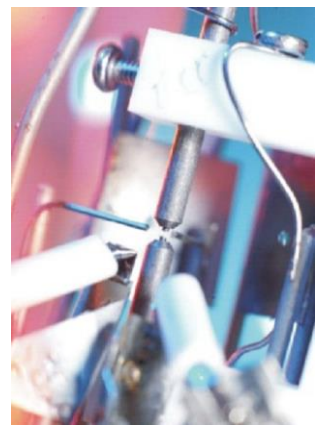

**Figure 7.** Photograph of an NPL endcap trap based on the design described in Ref.[14].

*5.2.2. Results and discussion.* At typical operating trap drive amplitude of 150 V at 14.55 MHz, the modelled temperature rise of the hottest part - the fine copper wires connecting the rf endcaps and dc electrodes - is ~2.8 K, see Figure 8. The temperature rise seen by the ion is $\Delta T_{ion(150V)} = (0.8\pm0.4)$ K. The four most important contributions to the uncertainty are the feedthrough temperature (0.20 K), voltage amplitude (0.20 K), emissivity of vacuum chamber inner surface (0.17 K) and Macor loss tangent (0.14 K). This effective temperature uncertainty corresponds to a trap induced BBR shift uncertainty of $(d\nu/\nu)_{E2} \sim 3\times10^{-18}$ and $(d\nu/\nu)_{E3} \sim 5\times10^{-19}$ for the $^{171}Yb^+$ E2 (436 nm) and $^{171}Yb^+$ E3 (467 nm) clock transitions, and $(d\nu/\nu)_{Sr} \sim 3\times10^{-18}$ for the $^{88}Sr^+$ E2 (674 nm) clock transition.

The material parameters used in simulation are listed in table 2 and table 5, and the resulting temperature distribution is presented in Figure 8 and corresponding heat balance in table 6.

**Table 5.** Parameters used in FEM simulation of Schrama design NPL endcap trap.

| Item | Insulators (between electrodes) | Holder, feed-through insulators, spacer | Window |
|---|---|---|---|
| Material | Alumina | Macor | Glass |
| Relative permittivity | 9.5 | 5.67 | 5.5 |
| Dielectric loss tangent | $1.0 \times 10^{-4}$ | $2.0\times10^{-3}$ | 0 |
| Thermal conductivity (W m$^{-1}$K$^{-1}$) | 35.4 | 1.46 | 1.4 |
| Emissivity | 0.855 | 0.92 | 0.2 |

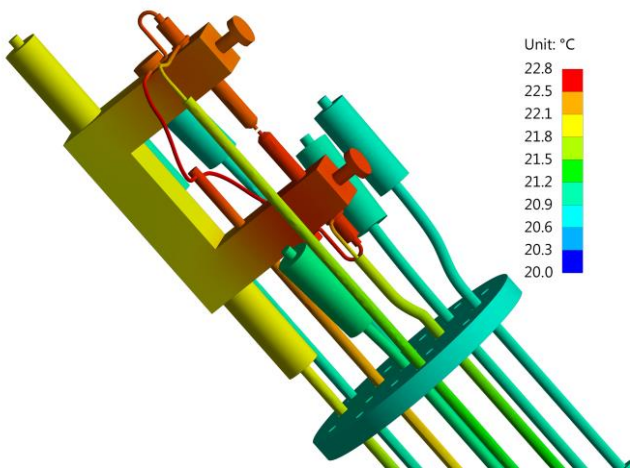

**Figure 8.** Modelled temperature distribution in NPL endcap trap for amplitude 150 V @ 14.55 MHz.

**Table 6.** Heat balance table for Schrama design NPL endcap trap for amplitude 150 V @ 14.55 MHz.

| | Part | Power (mW) |
|---|---|---|
| RF Heating | Macor holder | 9.9 |
| | Electrodes | 0.8 |
| | Alumina insulators | 2.5 |
| | Wires & Rods | 8.8 |
| | All other trap parts | 1.8 |
| | **Total heat generation** | **23.8** |
| Radiation | Macor holder | 11.9 |
| | Wires & Rods | 3.5 |
| | Electrodes | 0.6 |
| | All other trap parts | 3 |
| | **Total radiation** | **19** |
| Conduction | Rods | 5.7 |
| | **Total conduction** | **5.7** |

The existence of small vacuum gaps between the alumina insulators and the endcaps can cause discrepancies between the FEM simulation and the experimental realization of the trap. Since the relative permittivity of alumina is high ($\varepsilon_r$~9.5), a vacuum gap with relative permittivity $\varepsilon_r$=1 effectively reduces the capacitance of the trap, and consequently reduces currents and thus heat generation in the conductors. The wire which forms the inner endcaps sits within an alumina tube of slightly larger internal diameter, resulting in an uneven separation between the parts along the length of the tube. In the model a uniform spacing was assumed, and the separations of the insulator to rf endcap electrodes, and insulator to dc shield electrodes, were fitted to reach agreement between measured and simulated capacitances (without the gap the ion would see a temperature rise of ~1.2 K and the temperature rise at the hottest part of the trap would be ~4.5 K). The temperature distribution was experimentally measured using an IR camera and calibrated thermistors for a dummy trap of the same design as the operational traps. The highest values of temperature rise of (2.5±0.5) K were measured at the outer electrodes and the bottom part of the Macor holder, the temperature distribution also agrees well with the model (within ±0.15 K).

### 5.3. New NPL design endcap trap

*5.3.1. Trap description.* The new NPL design endcap trap [27] is designed to reduce heating in the trap structure, and thus reduce the necessity of precision modelling of the ion's BBR environment. It is intended for use for the $^{171}Yb^+$ and $^{88}Sr^+$ optical clocks, however the design is applicable for any atomic species.

The trap is supported by an OFHC copper 'C' which acts as the rf feed and thermal sink. Molybdenum rf and dc electrodes are separated by fused silica spacers and secured with 99.9% purity alumina bolts. The faces of the rf electrodes - which form the dominant solid angle subtended by the ion - are highly polished to a mean surface roughness of <20 nm to reduce the emissivity to $\varepsilon$ <0.02. The endcaps therefore act as mirrors and have nearly the same apparent temperature as the vacuum chamber and optical viewports.

In order to achieve high trap efficiencies and a pure quadrupole potential it is desirable to minimize the radial separation between the rf and dc electrodes in the trapping region to around 100 μm. At operating voltages of up to 1 kV this creates an electric field of $10^7$ V m$^{-1}$. As the dielectric heating depends of the square of the electric field, if the gap was filled by the dielectric spacer it would result in significant power dissipation and temperature rise. In this new design the dielectrics are laterally displaced from the trap centre into a region where the rf-dc separation can be increased to 2 mm without disturbing the trap, thus significantly reducing the electric field experienced by the dielectric. The small remaining amount of power that is dissipated into the spacer can then be removed from the vacuum system via the large thermal conductivity of the copper rf feedthrough (as discussed in section 2). The radial secular frequency $\omega_r$ of a single ytterbium ion for rf-amplitude 900 V at a frequency of 21 MHz is $\omega_{r(171Yb+)} = 2\pi \times 1$ MHz in this trap.

*5.3.2. Results and discussion.* At the test frequency and voltage of 900 V at 21 MHz, the temperature rise of the hottest part of the trap, with ideal thermal contact conductivity and without heating of the ceramic isolator in the vacuum feedthrough, is ~0.33 K (at the fine molybdenum wires which connect the grounded outer endcaps to copper wires), while the temperature rise seen by the ion is $\Delta T_{ion(900V)}$ ~ 0.05 K (Figure 9).

The temperature rise of the operational trap was measured using a calibrated Cedip Silver thermal imaging camera when the trap was operated at 21 MHz with voltage amplitudes from 250 V to 1 kV. The measurement revealed that the temperature rise in the trap follows the expected quadratic dependence with rf voltage (5) however at much higher levels than found in the idealized model. The origin of this temperature rise appears to be heat generated in the feedthrough insulator that propagates to the trap through the highly thermally conductive copper mount. This results in the entire trap having a nearly homogenous thermal distribution – equivalent to the idealized simulation – except with a constant offset from room temperature.

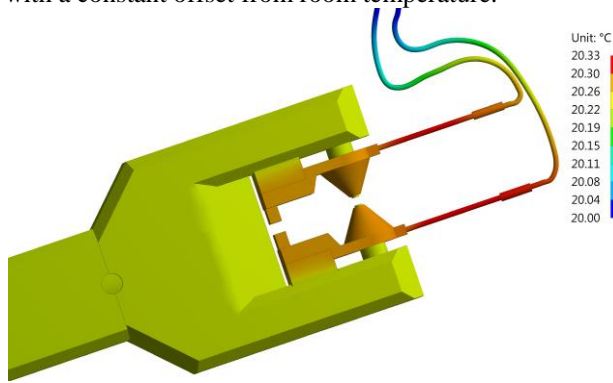

**Figure 9.** Modelled temperature distribution for the surface of the NPL design endcap trap for an applied rf potential of 900 V amplitude at 21 MHz (with feedthrough temperature fixed to ambient temperature 20°C).

As the construction of the feedthrough interior and ceramic-to-metal joins are not perfectly known the feedthrough temperature in the FEM model is fitted to the experimentally observed temperature rise of 6.5 K (Figure 10) to give $\Delta T_{ion(900V)} = (0.6 \pm 0.6)$ K. The four most important contributions to the uncertainty are the emissivity of molybdenum (0.47 K), emissivity of "ion" faces of molybdenum electrodes (0.26 K), uncertainty and homogeneity of vacuum chamber temperature (0.20 K) and voltage amplitude (0.13 K).

This effective temperature uncertainty corresponds to a trap induced BBR shift uncertainty of $(d\nu/\nu)_{E2}$ ~5×10$^{-18}$ and $(\delta\nu/\nu)_{E3}$ ~8×10$^{-19}$ for the $^{171}$Yb$^+$ E2 (436 nm) and $^{171}$Yb$^+$ E3 (467 nm) clock transitions respectively, and $(\delta\nu/\nu)_{Sr}$ ~4×10$^{-18}$ for the $^{88}$Sr$^+$ E2 (674 nm) clock transition. The material parameters used in simulation are listed in table 2 and table 7, the resulting temperature distribution is presented in Figure 10 and corresponding heat balance in table 8.

**Table 7.** Parameters used in FEM simulation of new NPL endcap trap.

| Item | Ceramic Bolt | Spacers, Window 1 | Window2 |
|---|---|---|---|
| Material | Alumina 99.9% | Fused silica | MgF2 |
| Relative permittivity | 10 | 3.82 | 5 |
| Dielectric loss tangent | 1.0×10$^{-4}$ | 1.5×10$^{-5}$ | 0 |
| Thermal conductivity (W m$^{-1}$K$^{-1}$) | 30 | 1.35 | 11.6 |
| Emissivity | 0.9 | 0.8 | 0.8 [a] |

[a] This is a "technical value" - because ANSYS software does not have radiation implemented for transparent bodies and to simulate the reflectivity of the window equal 0.2 requires to use emissivity equal 0.8 in the software. This value is used because this window (transparent in IR) is shielded by black screen at 20°C at atmosphere side during clock operation

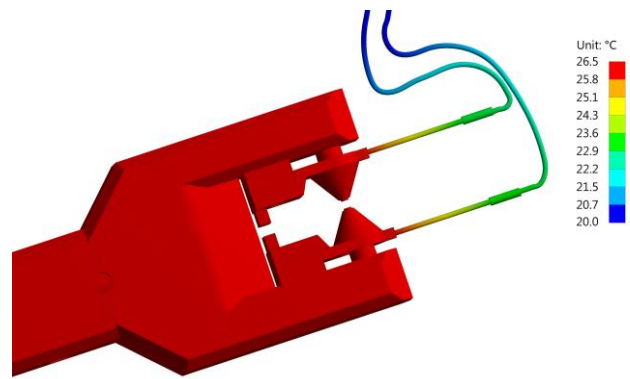

**Figure 10.** Modelled temperature distribution for the surface of the NPL design endcap trap for applied rf amplitude of 900 V @ 21 MHz with the boundary condition set to measured pin and feedthrough temperatures.

If the problem of the temperature rise in the feedthrough is solved, for example by heat-sinking the atmosphere-side of the feedthrough or by replacing the feedthrough with one which would generate negligible heat, then the temperature rise of the hottest part (the fine molybdenum wires which connect the GND endcaps to copper wires) would be ~0.15 K, while the temperature rise seen by the ion would be $\Delta T_{\text{ion}(900V)}$ ~0.05 K. This effective temperature uncertainty would correspond to a trap induced BBR shift of $(\delta\nu/\nu)_{E2}$ ~4×10$^{-19}$ and $(\delta\nu/\nu)_{E3}$ ~7×10$^{-20}$ for the $^{171}$Yb$^+$ E2 (436 nm) and $^{171}$Yb$^+$ E3 (467 nm) clock transitions respectively, and $(\delta\nu/\nu)_{Sr}$ ~4×10$^{-19}$ for the $^{88}$Sr$^+$ E2 (674 nm) clock transition.

**Table 8.** Heat balance for new NPL design endcap trap for amplitude 900 V @ 21 MHz.

|  | Part | Power (mW) |
|---|---|---|
| RF Heating | Ceramic bolts | 3.7 |
|  | Spacers | 1.0 |
|  | Molybdenum wires | 1.4 |
|  | Copper wires | 1.9 |
|  | All other trap parts | 0.5 |
|  | **Total heat generation** | **8.5** |
| Radiation | Ceramic bolts | 1.5 |
|  | Spacers | 2.4 |
|  | Endcaps(RF and GND) | 1.6 |
|  | Copper fingers (trap body) | 7.0 |
|  | All other trap parts | 0.8 |
|  | **Total radiation** | **13.4** |
| Conduction | Thick copper rod (Trap holder) | -23.2[a] |
|  | Thin copper wires | 18.4 |
|  | **Total conduction** | **-4.9** |

[a] Minus sign means that the heat flows to the trap (heat is generated outside the trap)

## 5.4. The PTB blade trap

*5.4.1. Trap description.* The linear PTB blade trap (Figure 11) is employed in the first generation of $^{27}$Al$^+$ ion quantum logic clock being developed at PTB [28]. Its electrode geometry is based on the Innsbruck linear trap design [29], but it is made from different materials to improve its thermal properties.

The trap consists of 4 blades, two hollow tip electrodes and 4 compensation electrodes (all titanium alloy TiAl6V4) mounted in 2 sapphire discs. The blades are connected to the rf vacuum feedthrough through 0.1 mm thick and 10 mm wide copper strips to improve conductivity, whereas 0.25 mm diameter copper wires are used for connecting the dc voltages. The distance between the tip electrodes is 5 mm and the distance between the rf electrodes and the trap centre is 0.8 mm.

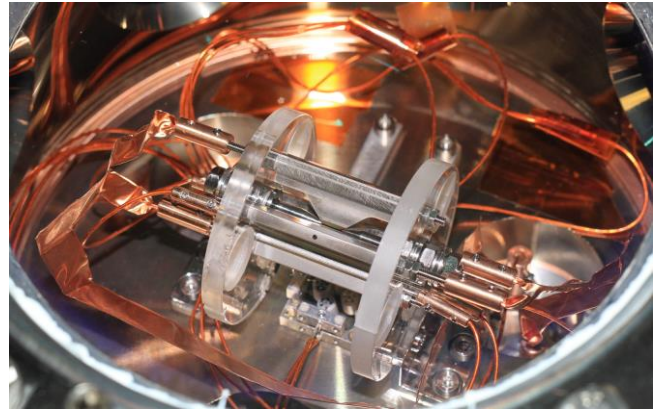

**Figure 11.** Linear trap – PTB blade design.

During operation the blade electrodes providing radial confinement are symmetrically driven at 24 MHz with drive voltage amplitude of about 350 V for a radial trapping frequency of 1.9 MHz for single calcium ions. From the thermal point of view, the mounting of the sapphire discs to the L-shaped brackets (made from an Al-Mg alloy, ENAW-6060) used to fix them to the baseplate of the vacuum chamber is important. One disc is mounted firmly with two bolts but the second disc is held using a sliding pivot to avoid stress from thermal expansion. The sapphire disc is not in contact with the bracket and the thermal contact with the pivot is poor as well.

A Pt100 temperature sensor was attached with vacuum compatible epoxy to the upper part of the sapphire disc mounted to the sliding pivot.

The radial secular frequency $\omega_r$ of aluminium ion for rf-amplitude 353 V at a frequency of 24 MHz is $\omega_{r(27Al+)} = 2\pi \times 2.8$ MHz in this trap.

*5.4.2. Results and discussion.* The thermal contact conductivities between different materials were set to 100 W m$^{-2}$·K$^{-1}$. This rather small value has been chosen by comparing the simulated temperature rise of the sapphire discs with the measured temperature rise 2.7 K at the Pt100 sensor. It is a plausible choice since no measures were taken to prepare surfaces for optimum thermal contact between different materials.

The material parameters used in simulation are listed in table 2 and table 9, the resulting temperature distribution is presented in Figure 12 and corresponding heat balance in table 10.

**Table 9.** Parameters used in FEM simulation of PTB linear Al$^+$ trap.

| Item | Connector insulator | Windows | Discs |
|---|---|---|---|
| Material | Alumina | Fused Silica | Sapphire |
| Relative permittivity | 9.5 | 3.82 | 10.4 |
| Dielectric loss tangent | 2×10$^{-4}$ | 1.5×10$^{-5}$ | 1.0×10$^{-4}$ |
| Thermal conductivity (W m$^{-1}$K$^{-1}$) | 35.4 | 1.35 | 24 |
| Emissivity | NA | 0.75 | 0.47 |

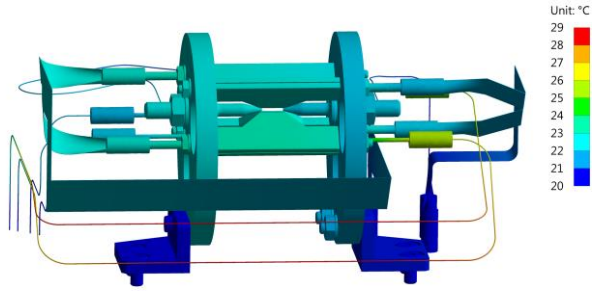

**Figure 12.** Modelled temperature distribution for symmetric drive amplitude 353 V @ 24 MHz.

The sapphire disc mounted by a pivot to the mounting brackets shows a temperature rise of 2.7 K which is 0.7 K more than the temperature rise of the disc mounted by bolts. This temperature rise is substantially smaller than that reported for a similar trap made from non-magnetic steel and Macor discs [30,31]

**Table 10.** Heat balance table for PTB linear Al$^+$ trap for symmetric drive 2 x amplitude 353 V @ 24 MHz.

| | part | Power (mW) |
|---|---|---|
| RF Heating | Sapphire discs | 32.4 |
| | Copper strips | 44.9 |
| | Blades and endcaps | 38.8 |
| | Wires | 11.0 |
| | All the remaining items within vac. chamber | 13.9 |
| | **Total heat generation** | **141** |
| Radiation | Sapphire discs | 39.6 |
| | Copper strips | 7.4 |
| | Blades and endcaps | 8.3 |
| | Wires | 1.5 |
| | All the remaining items within vac. chamber | 4.2 |
| | **Total radiation** | **61** |
| Conduction | RF connectors | 47.7 |
| | Mount (fastened by bolts with disc) | 22.1 |
| | All the remaining items within vac. chamber | 8.9 |
| | **Total conduction** | **78.7** |

The hottest parts of the trap are the fine copper wires leading to the compensation electrodes, whose temperature rises by 8.9 K. Including all trap components and the surrounding vacuum chamber, we deduce an effective increase of the BBR temperature at the ion's location of $\Delta T_{\text{ion}(353\text{V})} = (1.0\pm0.5)$ K. The four most important contributions to the uncertainty are the thermal conductivity of contact between Sapphire and ENAW6060 (0.33 K), uncertainty and homogeneity of vacuum chamber temperature (0.20 K), voltage amplitude (0.19 K) and sapphire loss tangent (0.17 K).

This effective temperature uncertainty corresponds to a fractional frequency uncertainty of $(\delta\nu/\nu)_{\text{Al}^+}\sim 3\times 10^{-20}$ for $^{27}$Al$^+$ [32]. This is well below the estimated uncertainty of other shifts for the Al$^+$ clock [33].

The temperature rise of this trap could be further reduced by improving the thermal conductance of the contacts. Especially the pivot joint between the sapphire disc and bolts is a very poor thermal conductor causing temperature increase of this disc. All thermal contacts could be improved by using a thin layer of a soft metal such as copper or indium between the contact areas. The contact through the pivot could be improved by mounting a flexible metal strip between the sapphire disc and the L-shaped mounting bracket.

### 5.5. PTB Segmented Chip Trap

*5.5.1. Trap description.* A scalable segmented trap design (Figure 13) was developed at PTB for the operation of an optical clock based on multiple ions [34],[35]. This design will be used for a clock based on $^{115}$In$^+$ ions sympathetically cooled by $^{172}$Yb$^+$ and for the 2nd generation of the $^{27}$Al$^+$ ion clock at PTB [36]. The trap is optimized for minimal axial micromotion and a prototype trap is in operation [35].

At an rf-amplitude of 1 kV and rf-frequency of 15.4 MHz, the radial secular frequencies $\omega_\text{r}$ of $^{172}$Yb$^+$, and $^{115}$In ions in this trap are $\omega_{\text{r}(172\text{Yb+})} = 2\pi\times1.0$ MHz and $\omega_{\text{r}(115\text{In+})} = 2\pi\times1.52$ MHz. For operation with $^{27}$Al$^+$, the rf amplitude would be considerably reduced, so that the temperature rise calculated here can be taken as a conservative upper bound.

The trap consists of four aluminium nitride (AlN) wafers (50 mm x 50 mm x 0.38 mm). The electrodes are formed by a sputtered and laser-structured gold layer of 4 μm thickness. Noise on the dc voltages is low-passed by on-board SMD filter electronics made of $Al_2O_3$ ceramics. Two Pt100 SMD temperature sensors are integrated on the two inner trap chips. To avoid heating by rf fields the sensors are protected by two capacitors placed in parallel. The AlN wafers are separated by 1 mm and 0.254 mm thick AlN spacers placed at the four corners. The trap stack is glued with a one component epoxy (Optocast 3410 Gen2) onto an AlN support board of 1 mm thickness and mounted by two AlN legs to the vacuum chamber.

*5.5.2. Results and discussion.* In a first step, the thermal conductivity between the trap chips was deduced by application of a known amount of heating power via a dc current through the Pt100 sensors, and measurement of the temperature distribution with the IR camera and thermistors attached to the trap mounts and the chamber. The comparison to FEM simulations showed that realistic thermal contact conductivities of about 1000 W m$^{-2}$ K$^{-1}$ are needed between the trap chips and spacers to match the experimental observations. If this value and the material parameters according to the datasheets (see table 2 and table 12) are used for modelling, the temperature rise of the hottest part of the trap is 1.45 K at the expected operating trap drive of 1 kV at 15.4 MHz. The temperature rise seen by the ion is $\Delta T_{ion(1kV)} = 0.51$ K. The measurement of the trap temperature by both IR camera and Pt100 sensors showed that the temperature rise of the hottest part of the trap (the rf electrodes and the AlN ceramics close-by) is 2.9 K for the same trap drive voltage and frequency. We attribute the observed higher temperature to a slightly higher loss tangent of the AlN substrates and to a contamination of the trap (see below). When the FEM simulations are performed with an adjusted AlN loss tangent of $8\times10^{-4}$ to match the experiment (Figure 14), the temperature rise at the location of the ion is $\Delta T_{ion(1kV)} = 1.0$ K. The maximum deviation between modelled and experimentally observed temperatures was 0.4 K. If the full temperature rise seen by the ion is taken as the temperature uncertainty, the uncertainty of the ion trap induced BBR shift is $(\delta\nu/\nu)_{Al+} \sim 5\times10^{-20}$ for $^{27}Al^+$ and $(\delta\nu/\nu)_{In+} \sim 2\times10^{-19}$ for $^{115}In^+$ [32]. Using the temperatures measured by the calibrated Pt100 sensors the uncertainty of the temperature rise seen by the ion is reduced to 0.1 K, which leads to a reduction of the BBR shift uncertainties by a factor of ten[9].

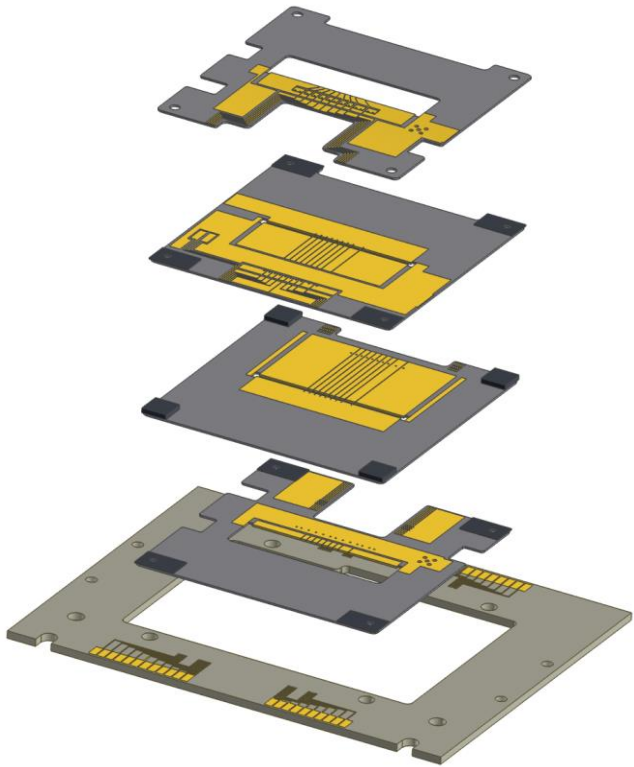

**Figure 13.** Exploded drawing of the chip-based segmented linear Paul trap.

A description of FEM parameters is given in table 2 and table 12 and heat balance in table 13.

---

[9] If the thermometers were not available, the estimated uncertainty of $\Delta T_{ion(1kV)}$ would be 0.4 K. The four most important contributions to the uncertainty would be the emissivity of sputtered gold (0.22 K), the voltage amplitude (0.20 K), uncertainty and homogeneity of vacuum chamber temperature (0.20 K) and the thermal conductivity of contact between AlN boards (0.16 K).

**Table 12.** Parameters used in FEM simulation of PTB's segmented chip trap

| Item | Boards, spacers, mounts |
|---|---|
| Material | AlN |
| Relative permittivity | 9 |
| Dielectric loss tangent | $3\times10^{-4}$ [a] |
| Thermal conductivity (W m$^{-1}$K$^{-1}$) | 160 |
| Emissivity | 0.73 |

[a] Value is taken from data sheet and is adjusted during simulations, see text.

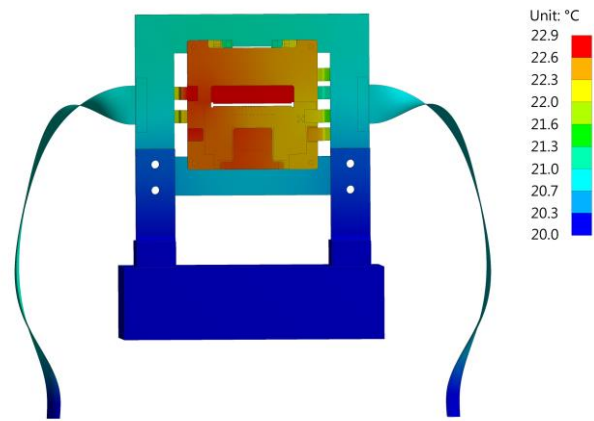

**Figure 14.** Modelled temperature distribution in the segmented chip trap for an rf amplitude of 1 kV at 15.4 MHz drive frequency. A thermal contact conductivity between the AlN boards and the spacers of 1000 W m$^{-2}$ K$^{-1}$ and an increased loss tangent $\tan(\delta) = 8\times10^{-4}$ were used.

**Table 13.** Heat balance table for PTB's segmented chip trap for amplitude 1000 V @ 15.4 MHz for an assumed loss tangent $\tan(\delta) = 8\times10^{-4}$.

| | Part | Power (mW) |
|---|---|---|
| RF Heating | Strips | 32.8 |
| | Boards | 204.7 |
| | Gold coatings | 38.1 |
| | Sum all others | 8.7 |
| | **Total heat generation** | **284.4** |
| Radiation | Strips | 3.7 |
| | Boards | 42.4 |
| | Gold coating | 1.6 |
| | Sum all others | 0.9 |
| | **Total radiation** | **48.6** |
| Conduction | Strips | 28.6 |
| | Mounts | 206.8 |
| | **Total conduction** | **235.3** |

For all simulations of this trap a simplified gold structure without SMD components was used. Heating due to onboard SMD electronics during rf operation was excluded by IR camera measurements on a dummy trap made of low heat conductivity material (0.6 W m$^{-1}$ K$^{-1}$) [35]

Measurements at different rf voltages have shown that the observed temperature rise follows the expected quadratic dependency (see (5)) only up to 500 V and increases faster for higher voltages. The observed additional heating which increases with increasing rf voltage cannot be attributed to higher rf losses in the AlN bulk, but is due to a contamination of the trap (deposition of a thin film nearby the rf electrode, probably due to some outgassing of heat-conductive paste used in the test vacuum chamber). The contamination could be removed by cleaning parts of the trap with acetone. From measurements with an rf voltage of 500 V we extrapolate that the temperature rise of the hottest part of the uncontaminated trap would be 2.0 K at a trap drive of 1 kV at 15.4 MHz, corresponding to a temperature rise seen by the ion of $\Delta T_{\text{ion(1kV)}}$ ~ 0.7 K. For the uncontaminated case simulations and experiments are matched by assuming a loss tangent of $\tan(\delta) = 5 \times 10^{-4}$.

Further improvements can be achieved, by electroplating thicker gold layers and using thicker copper strips to reduce resistive losses and to increase the thermal conductivity.

## 6. Conclusions

We have developed FEM models for several ion trap designs used in optical clocks and simulated the temperature rise of the blackbody radiation experienced by the ion under normal operational conditions. The simulated results were compared against thermal camera measurements of dummy traps and, in some cases, of actual operational traps. Initial discrepancies between simulations and measurements have been analyzed, explained, and corrected for either by improving thermal contact conductivities in the dummy traps or by adjusting the material parameters in the FEM models. Systematic comparisons between simulations and measurements until agreement is reached have revealed the importance of thermal contacts, manufacturing tolerances, heat generation in the rf vacuum feedthrough, the dielectric loss tangent of insulators (where batch-to-batch variations have been observed), using thick enough conductors, and vacuum chamber temperature homogeneity.

The BBR shift uncertainties corresponding to the temperature rise seen by the ions trapped in five traps of quite different design (Paul trap, endcap trap, linear blade trap, segmented chip trap) were evaluated. In all cases, the heating of the trap structure no longer makes limiting contributions in the clocks' uncertainty budgets, even for those traps that were constructed when low heating and thermal homogeneity were not design criteria of high priority. The BBR shift therefore does not present an obstacle to obtaining systematic uncertainties in the low 10$^{-18}$ range for optical clocks based on the Sr$^+$ and Yb$^+$ E2 transitions, and even lower for Yb$^+$ E3, In$^+$ and Al$^+$ because of the lower differential polarizabilities. Additional improvements can be obtained by optimized trap design as outlined here. Further reduction of BBR shift uncertainties would require better characterization of material parameters before the assembly of the trap combined with integration of temperature sensors in the traps (where it is possible by the trap design).

## Acknowledgments

This research is undertaken within the European Metrology Research Programme (EMRP) project SIB04 IonClock. The EMRP is jointly funded by the EMRP participating countries within EURAMET and the European Union. The CMI participation in the project is co-funded by the Ministry of Education, Youth and Sports of the Czech Republic (7Ax13011).